\documentclass[11pt]{article}
\usepackage[a4paper,lmargin=1in,rmargin=1in]{geometry}
\usepackage{amsmath}
\usepackage{amsfonts}
\usepackage{amsbsy}
\usepackage{amsthm}
\usepackage{tensor}
\usepackage{enumitem}
\newtheorem{thrm}{Theorem}[section]
\newtheorem{crlr}[thrm]{Corollary}
\newtheorem{lmm}[thrm]{Lemma}
\newtheorem{dfn}[thrm]{Definition}
\theoremstyle{definition}

\DeclareMathOperator{\Alt}{Alt}
\DeclareMathOperator{\sgn}{sgn}
\begin{document}

\title{Local differential calculus over Fedosov algebra}
\date{}
\author{Micha{\l} Dobrski
\footnote{michal.dobrski@knf.p.lodz.pl}
\\
\small
\emph{Center of Mathematics and Physics}
\\
\small
\emph{Technical University of {\L}\'od\'z,}
\\
\small
\emph{Al.~Politechniki 11, 90-924 {\L}\'od\'z, Poland}}
\maketitle
\abstract{In this paper the local differential calculus over Fedosov algebra is constructed using the trivialization isomorphism. The explicit formulas for deformed derivations are given. The resulting calculus can be used as a "building block" for a theory of Seiberg-Witten map with Fedosov type of noncommutativity.}
\section{Introduction}
The aim of this paper is to construct differential calculus over Fedosov algebra of formal power series with coefficients in functions on symplectic manifold. The term "differential calculus" refers here to an unital $\mathbb{N}$-graded algebra with nilpotent antiderivation i.e. to a structure analogous to the Cartan algebra of differential forms. 

The motivation for such investigation could be provided by the theory of Seiberg-Witten map (\cite{seibwitt}). The underlying noncommutativity of this theory is given by Moyal star product. One may ask about generalizations to other types of noncommutativity described by deformation quantization procedures e.g. to the Fedosov $*$-product on a symplectic manifold. When passing from Moyal to Fedosov product one encounters some difficulties. They  generally originate in the fact that operators $\frac{\partial}{\partial x^i}$ which are derivations of both undeformed algebra of functions and Moyal algebra are no longer derivations with respect to the Fedosov product. For this reason, the framework of the usual Cartan algebra cannot be used for developing consistent Seiberg-Witten map with Fedosov product. The construction presented in this paper may be regarded as a building block for such a theory.

The approach presented here is similar to that developed in \cite{asakawa1} and could be considered as its extension which enables explicit calculations (see concluding section for a discusion on relations between this paper and \cite{asakawa1}). Fedosov-type deformations of algebra of differential forms are also analyzed in \cite{dolgushev1} with methods of geometry of supermanifolds. However, the resulting deformation does not preserve $\mathbb{N}$-graded structure. The deformation of Cartan algebra with Moyal noncommutativity was considered in \cite{reuter}. In \cite{waldmann} global and general scheme for deformation of bimodule of sections of arbitrary vector bundle is presented. Unfortunately when trying to adopt these methods for the pourpose of the deformation of Cartan algebra, one faces severe problems with constructing compatible deformation of both tensor and wedge product.

The paper is organized as follows. First, the general scheme of Fedosov construction of the $*$-product on arbitrary symplectic manifold is recalled. Next (section 3) the trivialization procedure for Fedosov algebras (originally formulated in \cite{fedosov}) is analyzed. The explicit formulas (up to $h^2$) for trivialization isomorphism with arbitrary underlying homotopy of symplectic connections are given. The concept of trivialization turns out to be crucial for construction of differential calculus which is described in the fourth section. Finally, some concluding comments are given.

\section{Fedosov construction}
This section is given mainly for the pourpose of fixing the notations. Thus, the proofs are omitted, and the numbers of theorems in original formulation in $\cite{fedosov}$ are given. For detailed insight into geometrical ideas behind Fedosov construction one may refer to \cite{emmrwein}. Some further properties and examples could be found in \cite{tosiek}.

The Fedosov construction (\cite{fedosov0}, \cite{fedosov}) establishes a $*$-product on an arbitrary symplectic manifold $(\mathcal{M},\omega)$ with some fixed symplectic (torsionless and preserving $\omega$) connection $\partial$. The triple $(\mathcal{M}, \omega, \partial)$ is frequently called Fedosov manifold (\cite{gelfand}). The main idea of Fedosov procedure is to lift functions on manifold to "functions" on tangent bundle. These liftings can be multiplied using fibrewise Moyal product and it turns out that the result is also a lifting of some function (or more precisely -- of a formal power series with coefficients in functions) on the base manifold. 

In the first step, one constructs a bundle on the base manifold $\mathcal{M}$, called formal Weyl algebras bundle $W$, with fibres being algebras $W_x$ consisting of formal power series
\begin{equation}
\label{fedo_fps}
a(h,y)=\sum_{k,p \geq 0} h^k a_{i_1 \dots i_p}y^{i_1} \dots y^{i^p},
\end{equation}
where $y \in T_xM$, $a_{i_1 \dots i_p}$ are components of some symmetric covariant tensors in local Darboux coordinates and $h$ is a formal parameter. One prescribes degrees to monomials in formal sum (\ref{fedo_fps}) according to the rule
\begin{equation}
\deg (h^k a_{i_1 \dots i_p}y^{i_1} \dots y^{i^p} )=2k+p.
\end{equation}
For nonhomogeneous $a$ its degree is given by the lowest degree of nonzero monomials in formal sum (\ref{fedo_fps}). The operator $P_m$ extracts monomials of degree $m$ from given $a$
\begin{equation}
P_m(a)(h,y)=\sum_{2k+p=m} h^k a_{i_1 \dots i_p}y^{i_1} \dots y^{i^p}.
\end{equation}

The fiberwise $\circ$ product is defined by the Moyal formula
\begin{equation}
a \circ b = \sum_{m=0}^{\infty}\left( -\frac{ih}{2}\right)^m \frac{1}{m!}
\frac{\partial^m a}{\partial y^{i_1} \dots \partial y^{i_m}}
\omega^{i_1 j_1} \dots \omega^{i_m j_m}
\frac{\partial^m b}{\partial y^{j_1} \dots \partial y^{j_m}}.
\end{equation}
This definition is invariant under linear symplectomorphisms i.e. under transformations of $y^i$ generated by transitions between local Darboux coordinates on $\mathcal{M}$. 

We also consider bundle $W \otimes \Lambda$. Sections of this bundle can be locally written as
\begin{equation}
a=\sum h^k a_{i_1 \dots i_p j_1 \dots j_q}(x) y^{i_1} \dots y^{i^p} dx^{j_1} \wedge \dots \wedge dx^{j_q}.
\end{equation}
The $\circ$ product in $W \otimes \Lambda$ is defined by the rule $(a \otimes \eta) \circ (b \otimes \xi)=(a \circ b) \otimes (\eta \wedge \xi)$. The commutator of $a \in W \otimes \Lambda^r$ and $b \in W \otimes \Lambda^s$ is given by $[a,b]=a \circ b - (-1)^{rs} b \circ a$. It could be easily observed that the only elements which vanish on all commutators are scalar (containing no $y^i$) forms. The center generated by $[\,\cdot,\cdot\,]$ will be denoted as $Z$. The $Z$ valued $0$-forms are formal power series with coefficients in functions on $\mathcal{M}$ i.e. elements of $C^{\infty}(\mathcal{M})[[h]]$. 

One introduces an operator $\delta$ acting on elements of $W \otimes \Lambda$ as follows
\begin{equation}
\delta a=dx^k \wedge \frac{\partial a}{\partial y^k}=-\frac{i}{h}[\omega_{ij} y^i dx^j,a].
\end{equation}
Similarly, $\delta^{-1}$ acting on monomial $a_{km}$ with $k$-fold $y$ and $m$-fold $dx$ yields
\begin{equation}
\delta^{-1}a_{km}=\frac{1}{k+m}y^s \iota \left(\frac{\partial}{\partial x^s}\right) a_{km}
\end{equation}
for $k+m>0$ and $\delta^{-1}a_{00}=0$. Both $\delta$ and $\delta^{-1}$ are nilpotent and for $a \in W \otimes \Lambda^k$ the Leibniz rule $\delta(a \circ b)= (\delta a) \circ b + (-1)^k a \circ \delta b$ holds. An arbitrary  $a \in W \otimes \Lambda$ could be decomposed into
\begin{equation}
a=a_{00}+\delta \delta^{-1}a + \delta^{-1}\delta a.
\end{equation}

Let $a$ be a section of $W$. Symplectic connection $\partial$ can be extended to the Weyl bundle by the formula $\partial a=dx^i \wedge \partial_i a$, where $\partial_i a$ denotes covariant derivation of tensor fields in (\ref{fedo_fps}) with respect to $\frac{\partial}{\partial x^i}$. Using Darboux coordinates and connection coefficients $\tensor{\Gamma}{^l_{jl}}$ one may write $\partial a$ in the form
\begin{equation}
\partial a = da + \frac{i}{h}[1/2\Gamma_{ijk}y^i y^j dx^k,a],
\end{equation}
with $\Gamma_{ijk}=\omega_{il} \tensor{\Gamma}{^l_{jl}}$ (any further raising or lowering of indices is also performed by means of $\omega$).  When dealing with sections of $W \otimes \Lambda$, we can compute $\partial$ using the rule
\begin{equation}
\partial (\eta \circ a) = d\eta \circ a + (-1)^k \eta \circ \partial a,
\end{equation}
where $\eta$ is a scalar $k$-form. The $\circ$ Leibniz rule holds for $\partial$ and one could be interested in other connections with this property, namely in the connections of the form
\begin{equation}
\nabla=\partial + \frac{i}{h}[\gamma,\cdot\,],
\end{equation}
with $\gamma \in C^{\infty}(W \otimes \Lambda^1)$. One could calculate that $\nabla^2 = \frac{i}{h}[\Omega, \cdot \,]$ for curvature $2$-form $\Omega=R+\partial \gamma+\frac{i}{h}\gamma \circ \gamma$ with $R=\frac{1}{4}R_{ijkl}y^i y^j dx^k \wedge dx^l$ and $\tensor{R}{^i_{jkl}}$ -- the curvature tensor of symplectic connection on $\mathcal{M}$. In the case of symplectic connection in the Weyl bundle we obtain $\partial^{2}=\frac{i}{h}[R,\cdot\,]$. The connection $D$ is called Abelian if it is flat ($D^2=0$) i.e. if its curvature is a scalar form. The following theorem holds.
\begin{thrm}[Fedosov 5.2.2]
\label{fedo_abel}
For arbitrary symplectic connection $\partial$ there exists unique Abelian connection
\begin{equation}
D=-\delta+\partial+\frac{i}{h}[r,\cdot\,],
\end{equation}
satisfying the conditions: 
\begin{itemize}
\item $\Omega=-1/2\omega_{ij}dx^i \wedge dx^j$  is the corresponding curvature $2$-form,
\item $\delta^{-1}r=0$, 
\item $\deg r \geq 3$.
\end{itemize}
The $1$-form $r$ is the unique solution of the equation 
\begin{equation}
\label{fedo_abeliter}
r=\delta^{-1}R + \delta^{-1}\left( \partial r + \frac{i}{h}r \circ r\right).
\end{equation}
\end{thrm}
Section $a \in C^{\infty}(W)$ is called flat if $Da=0$. Flat sections form subalgebra of the algebra of all sections of $W$. We denote this subalgebra by $W_D$. If the underlying symplectic connection is flat and we work in Darboux coordinates for which $\Gamma_{ijk}=0$, then Abelian connection reads $D=d-\delta$. The corresponding subalgebra of flat sections is called \emph{trivial algebra} in this case.

For $a \in C^{\infty}(W \otimes \Lambda)$ define $Q(a)$ as a solution of the equation
\begin{equation}
b=a + \delta^{-1}(D+\delta)b
\end{equation}
with respect to $b$. One can prove that this solution is unique, and that $Q$ is linear bijection. Clearly $Q^{-1}a=a-\delta^{-1}(D+\delta)a$. The following theorem enables construction of $*$-product on the base manifold $\mathcal{M}$.
\begin{thrm}[Fedosov 5.2.4]
\label{fedo_qbiject}
The mapping $Q$ establishes bijection between $C^{\infty}(\mathcal{M})[[h]]$ and $W_D$.
\end{thrm}
Notice that for $a \in W_D$ one obtains $Q^{-1}(a)=a_0$, where $a_0$ is a central part of $a$. For $f,g \in C^{\infty}(\mathcal{M})[[h]]$ the $*$-product is defined according to the rule
\begin{equation}
f*g=Q^{-1}(Q(f) \circ Q(g)).
\end{equation}
The $*$-product obtained in this way is a globally defined associative deformation of the usual product of functions on $\mathcal{M}$. It also fulfills condition of locality (compare \cite{fedosov} for details) and the correspondence principle
($f*g-g*f=\frac{i}{h}\{f,g\}_P$, where $\{\,\cdot,\cdot \,\}_P$ is Poisson bracket induced by $\omega$).

In proofs of theorems \ref{fedo_abel} and \ref{fedo_qbiject} the \emph{iteration method} is used. Given equation of the form 
\begin{equation}
\label{fedo_itermet}
a=b +K(a)
\end{equation} 
one can try to solve it iteratively with respect to $a$, by putting $a^{(0)}=b$ and $a^{(n)}=b+K(a^{(n-1)})$. If $K$ is linear and \emph{raises degrees} (i.e. $\deg a < \deg K(a)$ or $K(a)=0$) then it could be easily deduced that the unique solution of (\ref{fedo_itermet}) is given by the series of relations $P_m(a)=P_m(a^{(m)})$. In the case of nonlinear $K$ (as in (\ref{fedo_abeliter})) the more careful analysis is required (\cite{fedosov}).

We need one more theorem for further pourposes.
\begin{thrm}[Fedosov 5.2.6]
\label{triv_deqlemma}
Equation $Da=b$ (for some given $b \in C^{\infty} (W \otimes \Lambda^p)$, $p>0$) has a solution if and only if $Db=0$. The solution may be chosen in the form $a=-Q \delta^{-1}b$.
\end{thrm}

\section{Trivialization}
In this section general methods developed by Fedosov are applied to the specific case of deformation quantization of symplectic manifold. To make this section self-contained, proofs of theorems directly related to the trivialization procedure are included (they strictly follow those of \cite{fedosov}).

The term \emph{trivialization} refers to the procedure of establishing isomorphism between some given algebra $W_D$ and the trivial algebra. The construction of this isomorphism is based on the following theorem.
\begin{thrm}[Fedosov 5.4.3]
\label{triv_trhm_liouv}
Let $D_t = d + \frac{i}{h}[\gamma (t),\cdot \:]$ be a family of Abelian connections parameterized by $t \in [0,1]$, and let $H(t)$ be $t$-dependent section of $W$ (called Hamiltonian) satisfying the following conditions:
\begin{enumerate}
\item $D_t H(t) - \dot{\gamma}(t)$ is a scalar form,
\item $\mathrm{deg}(H(t)) \geq 3$.
\end{enumerate}
Then, equation
\begin{equation}
\label{triv_heis}
\frac{da}{dt}+\frac{i}{h}[H,a]=0
\end{equation}
has the unique solution $a(t)$ for any given $a(0) \in W \otimes \Lambda$ and the mapping $a(0) \mapsto a(t)$ is injective for any $t \in [0,1]$. Moreover, if $a(0) \in W_{D_0}$ then $a(t) \in W_{D_t}$.
\end{thrm}
\begin{proof}
Equation (\ref{triv_heis}) can be integrated to give
\begin{equation}
\label{triv_heis_int}
a(t)=a(0)-\frac{i}{h} \int_0^t [H(\tau),a(\tau)] d \tau.
\end{equation}
Operator $\frac{i}{h} \int_0^t [H(\tau), \cdot \:] d \tau$ is linear and raises degrees since $\mathrm{deg}(H(t)) \geq 3$. Consequently existence, uniqueness and injectivity are provided by the iteration method.

Let $a(t)$ be a solution of (\ref{triv_heis}), then $D_t a(t)$ is also a solution. Indeed
\begin{equation}
\frac{d}{dt} D_t a = d \dot{a} + \frac{i}{h}[\dot{\gamma},a] + \frac{i}{h}[\gamma,\dot{a}] = D_t \dot{a} + \frac{i}{h}[\dot{\gamma},a].
\end{equation}
Using (\ref{triv_heis}) one can substitute $\dot{a}$ and obtain
\begin{equation}
\label{triv_abel_heis}
\frac{d}{dt} D_t a = - D_t \frac{i}{h}[H,a] + \frac{i}{h}[\dot{\gamma},a] = 
- \frac{i}{h}[D_t H - \dot{\gamma},a] - \frac{i}{h}[H,D_t a] = - \frac{i}{h}[H,D_t a],
\end{equation}
because $D_t H - \dot{\gamma}$ is a scalar 1-form. If $D_0 a(0)=0$ then the uniqueness of solution of (\ref{triv_heis}) ensures that $D_t a(t)=0$.
\end{proof}

One can fix $t=t_0$ in (\ref{triv_heis_int}) and rewrite it as
\begin{equation}
a(0) = a(t_0) + \frac{i}{h} \int_0^{t_0} [H(\tau),a(\tau)] d \tau.
\end{equation}
Then for any given $a(t_0)$ iteration method allows us to reconstruct unique $a(0)$ such that $a(t_0)$ is a solution of (\ref{triv_heis_int}) at $t=t_0$. Thus $a(0) \mapsto a(t)$ is bijective mapping. Same procedure repeated for $D_{t_0}a(t_0)=0$ in integrated equation (\ref{triv_abel_heis}) yields that if $a(t_0) \in W_{D_{t_0}}$ then $a(0) \in W_{D_0}$.

The solution of (\ref{triv_heis}) can be written in the form $a(t)=U^{-1}(t)a(0)U(t)$, where $U(t)$ is given as an iterational solution of
\begin{equation}
U(t)=1+\frac{i}{h} \int_0^t U(\tau) \circ  H(\tau) d \tau.
\end{equation}
Having the solution in this form, one easily sees that $a(0)\circ b(0) \mapsto a(t) \circ b(t)$. 
\begin{crlr}
\label{triv_isomcrlr}
The mapping $a(0) \mapsto a(t)$ defined in theorem \ref{triv_trhm_liouv} is an isomorphism between $W_{D_0}$ and $W_{D_t}$.
\end{crlr}

We are interested in constructing isomorphism between $W_D$ and the trivial algebra. This requires establishing homotopy of Abelian connections and compatible Hamiltonian. From now $W_{D_0}$ will denote the trivial algebra.
\begin{thrm}[Fedosov 5.5.1]
\label{triv_trhm_triv}
Any algebra $W_D$ on Fedosov manifold $(M, \omega, \partial)$ is locally isomorphic to the trivial algebra $W_{D_0}$ on $\mathbb{R}^{n}$.
\end{thrm}
\begin{proof}
Let $\mathcal{O}$ be a neighborhood of some point $x_0 \in M$, for which Darboux coordinates $x^i$ may be chosen. The symplectic connection  generates unique Abelian connection $D=d + \frac{i}{h}[\omega_{ij} y^i dx^j + 1/2 \Gamma_{ijk}y^i y^j dx^k + r,\cdot \:]$, where $r=1/8 R_{ijkl} y^i y^j y^k dx^l+ \dots$ is 1-form obtained from iterational procedure (\ref{fedo_abeliter}).

Consider a local homotopy of symplectic connections $\boldsymbol{\partial}^{(t)}$ such that for connection coefficients we have  $\boldsymbol{\Gamma}_{ijk}(0)=0$ and $\boldsymbol{\Gamma}_{ijk}(1)=\Gamma_{ijk}$. It generates homotopy of local Abelian connections 
\begin{equation}
D_t=d + \frac{i}{h}[\omega_{ij} y^i dx^j + 1/2 \boldsymbol{\Gamma}_{ijk}(t)y^i y^j dx^k + \boldsymbol{r}(t),\cdot \:]=
d+\frac{i}{h}[\boldsymbol{\gamma}(t),\cdot \:],
\end{equation}
satisfying $D_1=D$ and $D_0=d-\delta$ (trivial Abelian connection). Notice, that these Abelian connections have constant curvature $\boldsymbol{\Omega}(t)=-1/2 \omega_{ij}dx^i \wedge dx^j$. 

We look for a Hamiltonian being a solution of the equation $D_t H(t)=\dot{\boldsymbol{\gamma}}(t)$. According to the theorem \ref{triv_deqlemma} one have to check condition $D_t \dot{\boldsymbol{\gamma}}(t)=0$. We obtain
\begin{equation}
D_t \dot{\boldsymbol{\gamma}}(t)=d\dot{\boldsymbol{\gamma}}(t) + \frac{i}{h}[\boldsymbol{\gamma}(t),\dot{\boldsymbol{\gamma}}(t)]=\frac{d}{dt} \left( d\boldsymbol{\gamma}(t) + \frac{i}{h} \boldsymbol{\gamma}^2(t) \right)=\dot{\boldsymbol{\Omega}}(t)=0,
\end{equation}
and the Hamiltonian may be written as $H(t)=-Q_t \delta^{-1}\dot{\boldsymbol{\gamma}}(t)$. Since $d/dt$ commutes with $\delta^{-1}$ and the standard normalizing condition for an Abelian connection is $\delta^{-1}r=0$, one obtains 
\begin{equation*}
H(t)=-\frac{1}{6} Q_t(\dot{\boldsymbol{\Gamma}}_{ijk}(t)y^i y^j y^k),
\end{equation*}
with $\mathrm{deg}(H(t))\geq 3$. Thus assumptions of theorem \ref{triv_trhm_liouv} are fulfilled. The mapping defined therein is the desired isomorphism between $W_{D_0}$ and $W_D$.
\end{proof}

To obtain its explicit form  we need explicit form of $H(t)$. Using iteration method one could calculate $H(t)$ up to the fifth degree
\begin{equation}
\label{triv_hamilt}
\begin{split}
H(t)=&-\frac{1}{6} \dot{\boldsymbol{\Gamma}}_{ijk}(t)y^i y^j y^k
-\frac{1}{24} \boldsymbol{\partial}^{(t)}_{i} \dot{\boldsymbol{\Gamma}}_{jkl}(t)y^i y^j y^k y^l
-\frac{1}{120} \boldsymbol{\partial}^{(t)}_{i} \boldsymbol{\partial}^{(t)}_{j} \dot{\boldsymbol{\Gamma}}_{klm}(t)y^i y^j y^k y^l y^m+\\
&-\frac{1}{80} \boldsymbol{R}_{ijpk}(t) \tensor{\dot{\boldsymbol{\Gamma}}}{^p_{lm}}(t)y^i y^j y^k y^l y^m
+\frac{1}{32}h^2 \boldsymbol{R}_{ijkl}(t) \tensor{\dot{\boldsymbol{\Gamma}}}{^{ijk}}(t) y^l+\dots
\end{split}
\end{equation}
The expression $\boldsymbol{\partial}^{(t)}_{i} \dot{\boldsymbol{\Gamma}}_{jkl}(t)$ is used here to shorten notations. To obtain its explicit form one should calculate the covariant derivative treating $\dot{\boldsymbol{\Gamma}}_{jkl}(t)$ as a tensorial covariant object.

Let $T^{-1}:W_{D_0} \to W_D$ denote isomorphism mentioned in corollary \ref{triv_isomcrlr}. Its inverse $T: W_D \to W_{D_0}$ is called \emph{local trivialization} of $W_D$. Using (\ref{triv_hamilt}) when iterating equation (\ref{triv_heis_int}) one can compute first terms of $T^{-1}$. They read
\begin{multline}
T^{-1}(Q_0(a_0))=Q(a_0)+\\
+h^2 Q\left(
\frac{1}{24} \omega^{ls} \frac{\partial a_0}{\partial x^s} \int_0^1 \frac{\partial\boldsymbol{\Gamma}_{ijk}(\tau)}{\partial x^l} \dot{\boldsymbol{\Gamma}}^{ijk}(\tau) d\tau + \frac{1}{16} \omega^{ls} \frac{\partial^2 a_0}{\partial x^s \partial x^k} \Gamma^{ijk} \Gamma_{ijl}
+\frac{1}{24}\frac{\partial^3 a_0}{\partial x^i \partial x^j \partial x^k} \Gamma^{ijk}
\right)+\dots
\end{multline} 
Isomorphism $T^{-1}$ depends on the choice of homotopy $\boldsymbol{\Gamma}(t)$. However for homotopies of the form $\boldsymbol{\Gamma}_{ijk}(t)=f(t)\Gamma_{ijk}$ ($f:[0,1] \to \mathbb{R}$, $f(0)=0$, $f(1)=1$), the result is independent of the choice of $f$. In this case $T^{-1}$ reads
\begin{multline}
\label{triv_trivexplicit}
T^{-1}(Q_0(a_0))=Q(a_0)+\\
+h^2 Q\left(
\frac{1}{48} \omega^{ls} \frac{\partial a_0}{\partial x^s}  \frac{\partial\Gamma_{ijk}}{\partial x^l} \Gamma^{ijk} + \frac{1}{16} \omega^{ls} \frac{\partial^2 a_0}{\partial x^s \partial x^k} \Gamma^{ijk} \Gamma_{ijl}
+\frac{1}{24}\frac{\partial^3 a_0}{\partial x^i \partial x^j \partial x^k} \Gamma^{ijk}
\right)+\dots
\end{multline} 
and conversely
\begin{multline}
T(Q(a_0))=Q_0(a_0)+\\
-h^2 Q_0\left(
\frac{1}{48} \omega^{ls} \frac{\partial a_0}{\partial x^s}  \frac{\partial\Gamma_{ijk}}{\partial x^l} \Gamma^{ijk} + \frac{1}{16} \omega^{ls} \frac{\partial^2 a_0}{\partial x^s \partial x^k} \Gamma^{ijk} \Gamma_{ijl}
+\frac{1}{24}\frac{\partial^3 a_0}{\partial x^i \partial x^j \partial x^k} \Gamma^{ijk}
\right)+\dots
\end{multline} 
The above form of trivialization isomorphism will be used in the next section.

\section{Differential calculus}
In this section we are going to construct a differential calculus based on noncommutative Fedosov algebra of formal series. We follow ideas of Madore and collaborators (\cite{madore1}, \cite{madore2}) and also make extensive use of standard approach to Cartan algebra presented in \cite{spivak}.

First, let us recall algebraical definition of differential calculus (\cite{dubvio}, \cite{madore2}).
\begin{dfn}
A complex, unital and associative algebra $\mathcal{K}$ with product $\wedge$ is called differential calculus over $\mathcal{K}^0$ if it is $\mathbb{N}$-graded
\begin{enumerate}
\item $\displaystyle \mathcal{K}=\bigoplus_{n \geq 0}\mathcal{K}^n$, 
\item $\mathcal{K}^k \wedge \mathcal{K}^l \subset \mathcal{K}^{k+l}$
\end{enumerate} 
and it is equipped with compatible nilpotent antiderivation $d: \mathcal{K} \to \mathcal{K}$	
\begin{enumerate}[resume]
\item $d \mathcal{K}^l \subset \mathcal{K}^{l+1}$,
\item $d(\eta \wedge \xi)=(d \eta) \wedge \xi + (-1)^l \eta \wedge d \xi$ \hspace{5pt} 
for arbitrary $\eta \in \mathcal{K}^l$ and $\xi \in \mathcal{K}$ ,
\item $d^2=0$.
\end{enumerate} 
\end{dfn}


Let $\mathcal{O}$ be a neighborhood of some point $x_0$ for which trivialization theorem holds. Let $\Lambda^0=\mathcal{A}$ be usual algebra of functions on $\mathcal{O}$ and  $\Lambda^0_*=\mathcal{A}_*$ -- algebra of formal series obtained by Fedosov deformation quantization procedure. The initial step in our construction involves some observations about set $Der(\mathcal{A}_*)$. First, due to noncommutativity of $\mathcal{A}_*$, there are elements of $Der(\mathcal{A}_*)$ which could be written in the inner form $X(f)=\frac{i}{h}[\lambda \stackrel{*}{,} f]$, for $\lambda \in \mathcal{A}_*$ and $[\,\cdot \stackrel{*}{,} \cdot\,]$ denoting commutator with respect to Fedosov product. Also, unlike $Der(\mathcal{A})$, the set $Der(\mathcal{A}_*)$ is not a $\mathcal{A}_*$-(bi)module. For this reason, the construction of $\Lambda^1_*$ must be slightly different from construction of $\Lambda^1$. 

The natural and straightforward approach to the problem of building differential calculus over noncommutative algebra is presented in \cite{madore1} and \cite{madore2}. The key concept is to choose set $\mathcal{X}=\{X_1,\dots, X_n\}$ of $n$ derivations $X_i \in Der(\mathcal{A}_*)$, which is analogous to the frame in the classical geometry. We will use derivations of the form $X_i=\frac{i}{h}[\lambda_i \stackrel{*}{,} \cdot \,]$. The $\lambda_i$s may be chosen as follows
\begin{equation}
\lambda_i:=\omega_{ij}Q^{-1}T^{-1}Q_0 x^j.
\end{equation}
Using (\ref{triv_trivexplicit}) one finds that
\begin{equation*}
\lambda_i=\omega_{ij}x^j-\frac{h^2}{48}\frac{\partial \Gamma_{jkl}}{\partial x^i}\Gamma^{jkl}+\dots
\end{equation*}
Derivation $X_i$ acting on $f \in \mathcal{A}_*$ yields
\begin{multline}
\label{diffc_defder}
X_i(f)=\frac{i}{h}[\lambda_i \stackrel{*}{,} f]=Q^{-1}T^{-1}\frac{\partial}{\partial y^i} \left( TQf \right)
=\frac{\partial f}{\partial x^i}
-h^2\left\{
\frac{1}{48} \omega^{ls} \frac{\partial f}{\partial x^s}  \frac{\partial}{\partial x^i}\left(\frac{\partial\Gamma_{mjk}}{\partial x^l} \Gamma^{mjk}\right) \right.\\
\left.
+ \frac{1}{16} \omega^{ls} \frac{\partial^2 f}{\partial x^s \partial x^k} \frac{\partial (\Gamma^{mjk} \Gamma_{mjl})}{\partial x^i} 
+\frac{1}{24}\frac{\partial^3 f}{\partial x^m \partial x^j \partial x^k} \frac{\partial \Gamma^{mjk}}{\partial x^i}
\right\}+\dots
\end{multline}
The most important properties of $\mathcal{X}$ are the consequence of the following lemma.
\begin{lmm}
\label{diffc_lambdacomm}
The commutation relations for $\lambda_i$ are given by
\begin{equation}
\frac{i}{h}[\lambda_i \stackrel{*}{,} \lambda_j]=-\omega_{ij}.
\end{equation}
\end{lmm}
\begin{proof}
The straightforward calculation yields
\begin{equation*}
\begin{split}
\frac{i}{h}[\lambda_i \stackrel{*}{,} \lambda_j]=&
\frac{i}{h}[\omega_{ik}Q^{-1}T^{-1}Q_0 x^k \stackrel{*}{,} \omega_{jl}Q^{-1}T^{-1}Q_0 x^l]=\\
=&\omega_{ik}\omega_{jl} \frac{i}{h}Q^{-1}T^{-1}[Q_0 x^k \stackrel{\circ}{,} Q_0 x^l]= 
\omega_{ik}\omega^{kl}\omega_{jl}=-\omega_{ij}
\end{split}
\end{equation*}
\end{proof}
\begin{crlr}
\label{diffc_dercomm}
$X_i X_j=X_j X_i$ for each $X_i, X_j \in \mathcal{X}$.
\end{crlr}
\begin{proof}
Using lemma \ref{diffc_lambdacomm} and the Jacobi identity one obtains for $f \in \mathcal{A}_*$
\begin{equation*}
X_i X_j f= -\frac{1}{h^2} [\lambda_i \stackrel{*}{,} [\lambda_j\stackrel{*}{,}f]]=
\frac{1}{h^2} [f \stackrel{*}{,} [\lambda_i\stackrel{*}{,}\lambda_j]]+
\frac{1}{h^2} [\lambda_j \stackrel{*}{,} [f\stackrel{*}{,}\lambda_i]]=
-\frac{1}{h^2} [\lambda_j \stackrel{*}{,} [\lambda_i\stackrel{*}{,}f]]=X_j X_i f,
\end{equation*}
\end{proof}

Let $\mathcal{T}_*^k(\mathcal{X})$ denote the vector space (over $\mathbb{C}$) of mappings from $\mathcal{X}^k$ ($k$-fold product $\mathcal{X} \times \mathcal{X} \times \dots \times \mathcal{X}$) to $\mathcal{A}_*$. $\mathcal{T}_*^k(\mathcal{X})$ has a natural structure of $\mathcal{A}_*$-bimodule given by the relations 
\begin{gather}
(f*\eta)(X_{i_1},\dots,X_{i_k})=f*\eta(X_{i_1},\dots,X_{i_k}),\\
(\eta*f)(X_{i_1},\dots,X_{i_k})=\eta(X_{i_1},\dots,X_{i_k})*f
\end{gather}
for  $f \in \mathcal{A}_*$, $\eta \in \mathcal{T}_*^k(\mathcal{X})$ and $X_{i_1},\dots,X_{i_k} \in \mathcal{X}$. 

For $T \in \mathcal{T}_*^k(\mathcal{X})$ and $S \in \mathcal{T}_*^l(\mathcal{X})$, the tensor product $T \otimes_* S \in \mathcal{T}_*^{k+l}(\mathcal{X})$ may be defined as 
\begin{equation*}
(T \otimes_* S)(X_{i_1},\dots,X_{i_{k+l}}):=T(X_{i_1},\dots,X_{i_{k}})*S(X_{i_{k+1}},\dots,X_{i_{k+l}}).
\end{equation*}
\begin{thrm}
\label{diffc_tenprodprop}
The product $\otimes_*$ has the following properties
\begin{gather}
(S_1+S_2) \otimes_* T=S_1 \otimes_* T + S_2 \otimes_* T,\\
T \otimes_* (S_1+S_2)=T \otimes_* S_1 + T \otimes_* S_2,\\
(f*S) \otimes_* T= f*(S \otimes_* T),\\
S \otimes_* (T*f)= (S \otimes_* T)*f,\\
(S*f) \otimes_* T= S \otimes_* (f*T),\\
(S \otimes_* T) \otimes_* U = S \otimes_* (T \otimes_* U),
\end{gather}
for $S_1, S_2, S, T, U$ belonging to some (not necessarily the same) $\mathcal{T}_*^k(\mathcal{X})$, and $f \in \mathcal{A}_*$. 
\end{thrm}
Proof is a straightforward consequence of properties of $\mathcal{A}_*$.

One may introduce the exterior derivative of $f \in \mathcal{A}_*$ as a mapping $d_*f \in \mathcal{T}_*^1(\mathcal{X})$ defined by 
\begin{equation}
d_*f(X_i)=X_i(f).
\end{equation}
It could be easily observed that $d_*$ fulfills the Leibniz rule 
\begin{equation}
d_*(f*g)=(d_*f)*g+f*d_*g.
\end{equation} 
Our choice of $\mathcal{X}$ enables us to introduce "coframe" $\Theta=\{\theta^1, \dots, \theta^n \}$ consisting of $\theta^j \in \mathcal{T}_*^1(\mathcal{X})$ defined by
\begin{equation}
\theta^j:=d_*(\omega^{jk}\lambda_k)=d_*(Q^{-1}T^{-1}Q_0 x^j)
\end{equation}
By lemma \ref{diffc_lambdacomm} we calculate
\begin{equation}
\label{diffc_thetaonX}
\theta^j(X_i)=X_i(\omega^{jk}\lambda_k)=-\omega^{jk} \omega_{ik}=\tensor{\delta}{^j_i}.
\end{equation}
As a consequence one infers that each $\theta^j$ commutes with an arbitrary $f \in \mathcal{A}_*$, i.e. 
\begin{equation}
f * \theta^j=\theta^j * f.
\end{equation}
Define $\mathcal{B}_k$ as a set of all $k$-fold products $\theta^{i_1} \otimes_* \dots \otimes_*\theta^{i_k}$ ($\mathcal{B}_1=\Theta$).
\begin{thrm} 
\label{diffc_bkbasislmm}
$\mathcal{B}_k$ freely generates $\mathcal{A}_*$-bimodule $\mathcal{T}_*^k(\mathcal{X})$.
\end{thrm}
\begin{proof}
For arbitrary $T \in \mathcal{T}_*^k(\mathcal{X})$ one has
\begin{equation}
\label{diffc_bkbasis}
T=T(X_{i_1},\dots,X_{i_k})*\theta^{i_1} \otimes_* \dots \otimes_* \theta^{i_k},
\end{equation}
and equation $r_{i_1 \dots i_k} * \theta^{i_1} \otimes_* \dots \otimes_* \theta^{i_k}=0$ evaluated on $(X_{j_1},\dots,X_{j_k})$ yields $r_{j_1 \dots j_k}=0$. 
\end{proof}
One concludes that $\mathcal{B}_k$ is $\mathcal{A}_*$-basis of $\mathcal{T}_*^k(\mathcal{X})$.

We put $\Lambda^1_*=\mathcal{T}_*^1(\mathcal{X})$. Properties of $\mathcal{X}$ provide that construction of $\Lambda^k_*$ for $k>1$ may follow usual construction of $\Lambda$. The approach presented here is based on the classical textbook \cite{spivak}. The omitted proofs are just identical to those in \cite{spivak}.

We call $\eta \in \mathcal{T}_*^k(\mathcal{X})$ alternating if 
\begin{equation*}
\eta(X_{i_1},\dots,X_{i_p},\dots,X_{i_q},\dots,X_{i_k})=-\eta(X_{i_1},\dots,X_{i_q},\dots,X_{i_p},\dots,X_{i_k})
\end{equation*} 
for arbitrary $1 \leq p < q \leq k$. The subset of $\mathcal{T}_*^k(\mathcal{X})$ consisting of all alternating $\eta \in \mathcal{T}_*^k(\mathcal{X})$ is a $\mathcal{A}_*$-subbimodule of $\mathcal{T}_*^k(\mathcal{X})$. We put this submodule to be $\Lambda^k_*$. If $i_q=i_p$ for some $q \neq p$ then $\eta(X_{i_1},\dots,X_{i_k})=0$. Hence $\Lambda^k_*$ vanish for $k>n$.

To define exterior product we need some projection form $\mathcal{T}_*^k(\mathcal{X})$ to $\Lambda^k_*$. It could be chosen in the standard way. For $T \in \mathcal{T}_*^k(\mathcal{X})$ let
\begin{equation}
\Alt(T)(X_{i_1},\dots,X_{i_k}):=\frac{1}{k!}\sum_{\sigma \in S_k} \sgn(\sigma) T(X_{i_{\sigma(1)}},\dots,X_{i_{\sigma(k)}}),
\end{equation}
$S_k$ being group of all permutations of $\{1,\dots,k\}$, $\sgn(\sigma)=1$ for even and $\sgn(\sigma)=-1$ for odd permutations.
\begin{thrm}
\label{diffc_altprop}
The $\Alt$ operation has the following properties
\begin{gather}
\Alt(T) \in \Lambda^k_*,\\
\Alt(f*T+S*g)=f*\Alt(T)+\Alt(S)*g,\\
\label{diffc_altproj}
\Alt(\eta)=\eta\\
\Alt(\Alt(T))=\Alt(T)
\end{gather}
for $T,S \in \mathcal{T}_*^k(\mathcal{X})$, $\eta \in \Lambda^k_*$ and $f,g \in \mathcal{A}_*$. 
\end{thrm}
The second relation could be easily obtained from definition of $\Alt$. The others are proven in \cite{spivak}.
For $\eta \in \Lambda^k_*$ and $\xi \in \Lambda^l_*$ the exterior product $\eta \wedge_* \xi \in \Lambda^{k+l}_*$ is defined as
\begin{equation}
\eta \wedge_* \xi:= \frac{(k+l)!}{k!l!}\Alt(\eta \otimes_* \xi).
\end{equation}
\begin{thrm}
\label{diffc_extprodprop}
The $\wedge_*$ product has the following properties
\begin{gather}
(\xi_1+\xi_2)\wedge_* \eta=\xi_1 \wedge_* \eta + \xi_2 \wedge_* \eta,\\
\eta \wedge_* (\xi_1+\xi_2)=\eta \wedge_* \xi_1 + \eta \wedge_* \xi_1,\\
(f*\eta) \wedge_* \xi = f*(\eta \wedge_* \xi),\\
\eta \wedge_* (\xi*f) = (\eta \wedge_* \xi)*f,\\
(\eta*f) \wedge_* \xi = \eta \wedge_* (f*\xi),\\
(\eta \wedge_* \xi) \wedge_* \zeta = \eta \wedge_* (\xi \wedge_* \zeta),
\end{gather}
for $\eta \in \Lambda^k_*$, $\xi, \xi_1,\xi_2 \in \Lambda^l_*$ and $f \in \mathcal{A}_*$.
\end{thrm}
All except for the last of these relations are simple consequences of theorems \ref{diffc_tenprodprop} and \ref{diffc_altprop}. For associativity the proof is more elaborated and could be performed within three steps (compare \cite{spivak}). First, one should observe that for 
$T \in \mathcal{T}_*^k(\mathcal{X})$, $S \in \mathcal{T}_*^l(\mathcal{X})$, $\Alt(S)=0$ relation 
$\Alt(T \otimes_* S)=0=\Alt(S \otimes_* T)$
holds. Then it could be inferred that 
\begin{equation}
\label{diffc_wedgeasalt}
\Alt(\Alt(\eta \otimes_* \xi)\otimes_* \zeta)=\Alt(\eta \otimes_* \xi \otimes_* \zeta)=\Alt(\eta \otimes_* \Alt(\xi\otimes_* \zeta))
\end{equation}
for $\eta, \xi, \zeta$ in some (not necessarily the same) $\Lambda^k_*$. Finally one obtains
\begin{equation}
(\eta \wedge_* \xi) \wedge_* \zeta = \frac{(k+l+m)!}{k!l!m!}\Alt(\eta \otimes_* \xi \otimes_* \zeta) =\eta \wedge_* (\xi \wedge_* \zeta),
\end{equation}
for $\eta \in \Lambda^k_*, \xi \in \Lambda^l_*, \zeta \in \Lambda^m_*$. Each of these steps is independent from (non)commutativity of underlying algebra (see \cite{spivak} for details). They are rather based on properties of groups $S_k$ and theorems  \ref{diffc_tenprodprop} and \ref{diffc_altprop}.

Theorem \ref{diffc_extprodprop} justifies extension of exterior product to $0$-forms. For $f\in \Lambda^0_*=\mathcal{A}_*$ and $\eta \in \Lambda^k_*$ we put $f \wedge_* \eta:=f * \eta$ and $\eta \wedge_* f:= \eta * f$.
 
Notice that in general one cannot obtain relation analogous to $\eta \wedge \xi=(-1)^{kl} \xi \wedge \eta$. Fortunately, due to (\ref{diffc_thetaonX}), the following formula holds
\begin{equation}
\label{diffc_thetacomm}
\theta^i \wedge_* \theta^j = -\theta^j \wedge_* \theta^i,
\end{equation}
for arbitrary $\theta^i, \theta^j \in \Theta$, and  in general 
\begin{equation}
\label{diffc_thetacommgen}
\theta^{i_1} \wedge_* \theta^{i_2} \wedge_* \dots \wedge_* \theta^{i_k} = \sgn (\sigma ) \theta^{i_{\sigma(1)}} \wedge_* \theta^{i_{\sigma(2)}} \wedge_* \dots \wedge_* \theta^{i_{\sigma(k)}},
\end{equation}
for $\sigma \in S_k$ and $\theta^{i_1}, \dots, \theta^{i_k} \in \Theta$.

Using lemma \ref{diffc_bkbasislmm} and formulas (\ref{diffc_altproj}), (\ref{diffc_wedgeasalt}) one could represent arbitrary $\eta \in \Lambda^k_*$ as
\begin{equation}
\eta=\frac{1}{k!} \eta(X_{i_1},\dots,X_{i_k})*\theta^{i_1} \wedge_* \dots \wedge_* \theta^{i_k}.
\end{equation}
Applying (\ref{diffc_thetacommgen}) one reduces the above relation to
\begin{equation}
\eta=\sum_{1 \leq i_1 < \dots <i_k \leq n} \eta(X_{i_1},\dots,X_{i_k})*\theta^{i_1} \wedge_* \dots \wedge_* \theta^{i_k}.
\end{equation}
The $\mathcal{A}_*$-linear independence of the set $\mathcal{C}_k:=\{\theta^{i_1} \wedge_* \dots \wedge_* \theta^{i_k} :  1 \leq i_1 < \dots <i_k \leq n\}$ could be proven. Hence, $\mathcal{C}_k$ is $\mathcal{A}_*$-basis of $\Lambda^k_*$. and \begin{equation}
\dim(\Lambda^k_*)=\binom{n}{k}.
\end{equation}
Moreover (\ref{diffc_thetacommgen}) guarantees that
\begin{equation}
\frac{1}{k!} \eta_{i_1 \dots i_k} *\theta^{i_1} \wedge_* \dots \wedge_* \theta^{i_k}=\frac{1}{k!} \eta_{[i_1 \dots i_k]} *\theta^{i_1} \wedge_* \dots \wedge_* \theta^{i_k}.
\end{equation} 
One could also infer that any $\eta \in \Lambda^k_*$ could be written as
\begin{equation}
\eta=\frac{1}{k!} \eta_{i_1 \dots i_k} *\theta^{i_1} \wedge_* \dots \wedge_* \theta^{i_k}
\end{equation}
in the unique manner, provided that $\eta_{i_1 \dots i_k}$ is totally antisymmetric. Notice, that given two forms $\eta=\frac{1}{k!} \eta_{i_1 \dots i_k}* \theta^{i_1} \wedge_* \dots \wedge_* \theta^{i_k}$ and $\xi=\frac{1}{l!} \xi_{j_1 \dots j_l}* \theta^{j_1} \wedge_* \dots \wedge_* \theta^{j_l}$ their exterior product may be written as
\begin{equation}
\eta \wedge_* \xi =\frac{1}{k! l!} \, \eta_{i_1 \dots i_k} * \xi_{j_1 \dots j_l} *\theta^{i_1} \wedge_* \dots \wedge_* \theta^{i_k} \wedge_* \theta^{j_1} \wedge_* \dots \wedge_* \theta^{j_l}.
\end{equation}

We are ready to extend $d_*$ to forms of higher degree. Define
\begin{equation}
d_* \left( \frac{1}{k!} \eta_{i_1 \dots i_k} *\theta^{i_1} \wedge_* \dots \wedge_* \theta^{i_k} \right):=
\frac{1}{k!} X_j(\eta_{i_1 \dots i_k})* \theta^{j} \wedge_* \theta^{i_1} \wedge_* \dots \wedge_* \theta^{i_k}.
\end{equation}
Suppose that $\frac{1}{k!} \eta_{i_1 \dots i_k} *\theta^{i_1} \wedge_* \dots \wedge_* \theta^{i_k}=\frac{1}{k!} \tilde{\eta}_{i_1 \dots i_k} *\theta^{i_1} \wedge_* \dots \wedge_* \theta^{i_k}$. Then $\eta_{[i_1 \dots i_k]}=\tilde{\eta}_{[i_1 \dots i_k]}$ and
\begin{multline}
d_* \left( \frac{1}{k!} \eta_{i_1 \dots i_k} *\theta^{i_1} \wedge_* \dots \wedge_* \theta^{i_k} \right)=
\frac{1}{k!} X_{[j}(\eta_{i_1 \dots i_k]}) *\theta^{j} \wedge_* \theta^{i_1} \wedge_* \dots \wedge_* \theta^{i_k}=\\
=\frac{1}{k!} \frac{1}{k+1}\left(X_{j}(\eta_{[i_1 \dots i_k]}) - X_{i_1}(\eta_{[j \dots i_k]}) - \dots - X_{i_k}(\eta_{[i_1 \dots j]})\right) *\theta^{j} \wedge_* \theta^{i_1} \wedge_* \dots \wedge_* \theta^{i_k}=\\
=\frac{1}{k!} \frac{1}{k+1}\left(X_{j}(\tilde{\eta}_{[i_1 \dots i_k]}) - X_{i_1}(\tilde{\eta}_{[j \dots i_k]}) - \dots - X_{i_k}(\tilde{\eta}_{[i_1 \dots j]})\right) *\theta^{j} \wedge_* \theta^{i_1} \wedge_* \dots \wedge_* \theta^{i_k}=\\
=\frac{1}{k!} X_{[j}(\tilde{\eta}_{i_1 \dots i_k]}) *\theta^{j} \wedge_* \theta^{i_1} \wedge_* \dots \wedge_* \theta^{i_k}=
d_* \left( \frac{1}{k!} \tilde{\eta}_{i_1 \dots i_k} *\theta^{i_1} \wedge_* \dots \wedge_* \theta^{i_k} \right),
\end{multline}
hence $d_*$ is well defined. Notice, that above definition of $d_*$ is compatible with definition of $d_*$ for $0$-forms since $d_*f=d_*f(X_j) *\theta^j=X_j(f) *\theta^j$ for $f \in \mathcal{A}_*$.
\begin{thrm} 
The $d_*$ operator has the following properties 
\begin{enumerate}
\item $d_* d_*=0$,
\item $d_* (\eta \wedge_* \xi)= (d_* \eta) \wedge_* \xi + (-1)^k \eta \wedge_* (d_* \xi)$ for $\eta \in \Lambda^k_*$ and $\xi \in \Lambda^l_*$.
\end{enumerate}
\end{thrm}
\begin{proof}
Using corollary \ref{diffc_dercomm} and formula (\ref{diffc_thetacomm}) one calculates
\begin{multline*}
d_* d_* \eta =  \frac{1}{k!} X_k(X_j(\eta_{i_1 \dots i_k})) *\theta^{k} \wedge_* \theta^{j} \wedge_* \theta^{i_1} \wedge_* \dots \wedge_* \theta^{i_k}=\\
=-\frac{1}{k!} X_j(X_k(\eta_{i_1 \dots i_k})) *\theta^{j} \wedge_* \theta^{k} \wedge_* \theta^{i_1} \wedge_* \dots \wedge_* \theta^{i_k}=-d_* d_* \eta=0.
\end{multline*}
For the exterior product 
\begin{multline*}
d_*(\eta \wedge_* \xi) =d_*\left(\frac{1}{k! l!} \, \eta_{i_1 \dots i_k} * \xi_{j_1 \dots j_l} *\theta^{i_1} \wedge_* \dots \wedge_* \theta^{i_k} \wedge_* \theta^{j_1} \wedge_* \dots \wedge_* \theta^{j_l} \right)=\\
=\frac{1}{k! l!} \, X_j(\eta_{i_1 \dots i_k}) * \xi_{j_1 \dots j_l} *\theta^{j} \wedge_* \theta^{i_1} \wedge_* \dots \wedge_* \theta^{i_k} \wedge_* \theta^{j_1} \wedge_* \dots \wedge_* \theta^{j_l}+\\
+\frac{1}{k! l!} \, \eta_{i_1 \dots i_k} * X_j(\xi_{j_1 \dots j_l}) *\theta^{j} \wedge_* \theta^{i_1} \wedge_* \dots \wedge_* \theta^{i_k} \wedge_* \theta^{j_1} \wedge_* \dots \wedge_* \theta^{j_l}=\\
\frac{1}{k! l!} \, X_j(\eta_{i_1 \dots i_k}) * \xi_{j_1 \dots j_l} *\theta^{j} \wedge_* \theta^{i_1} \wedge_* \dots \wedge_* \theta^{i_k} \wedge_* \theta^{j_1} \wedge_* \dots \wedge_* \theta^{j_l}+\\
+(-1)^k \frac{1}{k! l!} \, \eta_{i_1 \dots i_k} * X_j(\xi_{j_1 \dots j_l}) * \theta^{i_1} \wedge_* \dots \wedge_* \theta^{i_k} \wedge_* \theta^{j} \wedge_*\theta^{j_1} \wedge_* \dots \wedge_* \theta^{j_l}=\\
=(d_* \eta) \wedge_* \xi + (-1)^k \eta \wedge_* (d_* \xi).
\end{multline*}
\end{proof}
One concludes that $\Lambda_*=\Lambda_*^0 \oplus \cdots \oplus \Lambda_*^n$ together with $d_*$ is a differential calculus over $\mathcal{A}_*$.
\section{Final comments}
The main result of this paper is the explicit construction of local differential calculus over Fedosov algebra of formal power series with coefficients in functions on symplectic manifold. The approach presented here is generally inspired by some standard procedures of noncommutative differential geometry (\cite{madore1},\cite{madore2}). Quite similar analysis could be found in \cite{asakawa1}. The main difference is that in our approach we do not \emph{postulate} existence of $\lambda_i$s with commutation relations given by lemma \ref{diffc_lambdacomm}, but we rather \emph{construct} them using trivialization procedure. Thus, we are able to give some explicit formulas for deformed case e.g. (\ref{diffc_defder}). The important advantage of this method is that the resulting differential calculus could be regarded as the deformation of the usual one i.e. obtained corrections vanish either at $h=0$ or $\Gamma_{ijk} \equiv 0$. On the other hand, construction presented here is local (limited to some open subset for which trivialization theorem holds) and one could hardly point out some method for gluing together differential calculi from different Darboux coordinates. 
\section*{Acknowledgments}
I would like to thank professor Maciej Przanowski for suggesting topic of investigation, giving many helpful remarks and reviewing the initial version of this paper. I am also grateful to Jaromir Tosiek and Sebastian Forma\'{n}ski for a number of discussions on Fedosov construction and noncommutative geometry. 

\end{document}